\documentclass[fp]{jpsj3}
\usepackage{txfonts}
\usepackage[dvipdfm]{}
\def\tc{$T_c$} \def\tl{$1/T_1$} \def\kb{$k_BT_c$} 
\title{NMR and NQR Studies on \\ Non-centrosymmetric Superconductors Re$_7$B$_3$, LaBiPt, and BiPd}

\author{
\name{Kazuaki \surname{Matano}}$^1\thanks{E-mail: matano@psun.phys.okayama-u.ac.jp}$,
\name{Satoki \surname{Maeda}}$^1$, 
\name{Hiroki \surname{Sawaoka}}$^1$,
\name{Yuji  \surname{Muro}}$^{2}$\thanks{Present address: Department of Liberal Arts and Sciences, Toyama Prefectural University, Toyama 939-0398, Japan.},
\name{Toshiro \surname{Takabatake}}$^2$,
\name{Bhanu \surname{Joshi}}$^3$,
\name{Srinivasan \surname{Ramakrishnan}}$^3$,
\name{Kenji \surname{Kawashima}}$^4$,
\name{Jun \surname{Akimitsu}}$^4$,
\name{Guo-qing \surname{Zheng}}$^{1,5}$
}

\inst{
$^1$Department of Physics, Okayama University, Okayama 700-8530, Japan\\
$^2$Department of Quantum Matter, ADSM, Hiroshima University,\\
Higashihiroshima, Hiroshima 739-8530, Japan\\
$^3$Department of Condensed Matter Physics and Materials Science, \\ Tata Institute of Fundamental Research, Mumbai 400005, India \\
$^4$Department of Physics and Mathematics, Aoyama Gakuin University, Sagamihara 252-5258, Japan \\
$^5$Institute of Physics and Beijing National Laboratory for Condensed Matter
Physics,\\ Chinese Academy of Sciences, Beijing 100190, China\\
} 

\abst{
We report the nuclear magnetic resonance (NMR) and  nuclear quadrupole resonance (NQR) 
measurements for non-centrosymmetric superconductors Re$_7$B$_3$, LaBiPt, and BiPd containing heavy elements. 
For all three compounds, the spin-lattice relaxation rate \tl\ shows a  coherence peak just below \tc\ and
 decreases exponentially at low temperatures,
which indicates that   an isotropic superconducting gap is dominant in these compounds.  In BiPd, the height of the coherence peak  just below $T_c$ is much suppressed, which  suggests that there exists a substantial component  of gap with nodes in this compound.
Our results indicate that heavy element is  not the only factor, but the extent of inversion symmetry breaking is also important to 
induce a large spin-orbit coupling and an unconventional superconducting state. 
}
\kword{Non-centrosymmetric superconductors, NMR, NQR}

\begin{document}
\maketitle

\section{Introduction}
In recent years, non-centrosymmetric superconductors have attracted much interest. 
In non-centrosymmetric systems, an antisymmetric spin-orbit coupling (ASOC) interaction is induced and a parity-mixed superconducting state is allowed. The extent of the parity mixing is determined by the strength of the ASOC.
\cite{PhysRevLett.87.037004,PhysRevLett.92.097001,1367-2630-6-1-115} 
Moreover, the existence of topologically protected zero-energy surface- or edge-states
has been pointed out recently in non-centrosymmetric superconductors.
\cite{PhysRevLett.105.097002,RevModPhys.83.1057}

Since the discovery of the non-centrosymmetric compound CePt$_3$Si,\cite{PhysRevLett.92.027003} 
many superconductors without inversion symmetry have been reported.
They can be categorized into two types. Namely, the strongly-correlated electron systems such as UIr\cite{UIr} and
CeRh(Ir)Si$_3$,\cite{CeRhSi3,CeIrSi3} and the weakly correlated electron systems that include Li$_2$Pd$_3$B, Li$_2$Pt$_3$B
and Mg$_{10}$Ir$_{19}$B.\cite{Li2Pd3B,Li2Pt3B,Mg10Ir19B16}
In the former class of materials, the electron correlations seem to play an important role in determining the superconducting properties.
The latter class  is therefore more suitable for studying the pure effects of inversion-symmetry breaking
and ASOC interaction.


The isostructural Li$_2$Pd$_3$B and Li$_2$Pt$_3$B show considerable differences.
\cite{Nishiyama_PhysRevB.71.220505,Nishiyama_PhysRevLett.98.047002,Harada_PhysRevB.86.220502}
The $^{11}$B spin-lattice relaxation rate (\tl) in Li$_2$Pd$_3$B shows a coherence peak just below $T_c$ and decreases exponentially
at low temperatures.\cite{Nishiyama_PhysRevB.71.220505}
On the contrary, \tl\ of Li$_2$Pt$_3$B decreases below $T_c$ without a coherence peak and follows $T^3$ variation.\cite{Nishiyama_PhysRevLett.98.047002} 
Also, the Knight shift ($K$)   changes below \tc\ in Li$_2$Pd$_3$B but does not change across \tc\ in the case of Li$_2$Pt$_3$B.
These results suggest isotropic gap, spin singlet superconductivity in Li$_2$Pd$_3$B, 
but nodal gap, spin triplet superconductivity in Li$_2$Pt$_3$B.
In this case, a strikingly different ASOC was believed to be the origin for the different superconducting state. 
The strength of the ASOC interaction is proportional to the square of the atomic number $Z^2$. Since Pt (5d) is below Pd (4d) in the periodic table of  elements,  the ASOC interaction is much stronger 
in Li$_2$Pt$_3$B  than Li$_2$Pd$_3$B.


In order to shed more light on the subject, studies on other non-centrosymmetric superconductors containing heavy elements are desired.
In this paper, we report the nuclear magnetic resonance (NMR) and nuclear quadrupole resonance (NQR) 
studies on non-centrosymmetric superconductors Re$_7$B$_3$, LaBiPt, and BiPd that contain heavy elements.

Re$_7$B$_3$ ($T_c$$\sim$3.3 K) has a hexagonal unit cell with the space group of $P6_3mc$.\cite{Re7B3}
LaBiPt ($T_c$$\sim$0.9 K) is one of the half-Heusler compounds with the space group of $F\bar{4}3m$ and 
is proposed to be a candidate for a topological superconductor.\cite{LaBiPt,nature-mat-9-546,PhysRevLett.105.096404}
The isostructural compounds YBiPt and LuBiPt are also reported to be  superconductive.\cite{PhysRevB.84.220504,2013arXiv1302.1943T}
BiPd ($T_c$$\sim$3.8 K) has a monoclinic structure,\cite{BiPd} the same structure with UIr,\cite{UIr}  with the space group of $P2_1$.
From the point contact Andreev reflection measurements, multiple superconducting gaps were suggested in BiPd.\cite{PhysRevB.86.094520}

These three compounds contain the heavy elements Re ($Z$=75), Bi ($Z$=83) and Pt ($Z$=78).
Therefore, we expect a strong ASOC interaction and novel superconductivity. 
Moreover, these compounds have  non-centrosymmetric crystal structures different from each other.
Studying the superconducting properties, we aim to understand the relationship between superconductivity and  the ASOC with different  crystal structures.

\section{Experimental}
The samples of Re$_7$B$_3$ were prepared by mixing the appropriate amounts of powders Re (99.99\%) and amorphous Boron (B) 
(99\%) in a dry box and synthesized by arc melting in high purity Ar gas.\cite{Re7B3}
The sample of LaBiPt was made by use of the Bridgeman technique in hermetically sealed Mo crucibles.\cite{LaBiPt}
The sample of BiPd was made by heating the individual components (Bi, 99.999\% pure, and Pd, 99.99\% pure) 
in a high-purity Alumina crucible with a pointed bottom, which is kept in a quartz tube that is sealed under a vacuum of 10$^{-6}$ mbar. 
Initially, the contents were heated up to 650 $^\circ$C (melting point of BiPd) in 12 h and then kept at 650 $^\circ$C for 12 h. Thereafter, 
it was slowly cooled to 590 $^\circ$C with a rate of 1 $^\circ$C/h and, finally, the furnace was switched off.\cite{BiPd}

For NMR/NQR measurements, the samples were crushed into powders. 
The $T_c$ at zero and a finite magnetic field $H$ was determined 
by measuring the ac susceptibility using the in situ NMR/NQR coil.
The $T_c$ of LaBiPt, and BiPd is 3.8 K and 1.2 K, respectively.
The \tc\ of Re$_7$B$_3$ is 3.3 K at zero filed and 2.6 K at 0.24 T.

A standard phase-coherent pulsed NMR spectrometer was used to collect data.
Measurements below 1.4 K were carried out in a $^3$He-$^4$He dilution refrigerator.
\section{Results and Discussion}
\begin{figure}[htbp]
\begin{center}\includegraphics[clip,width=70mm]{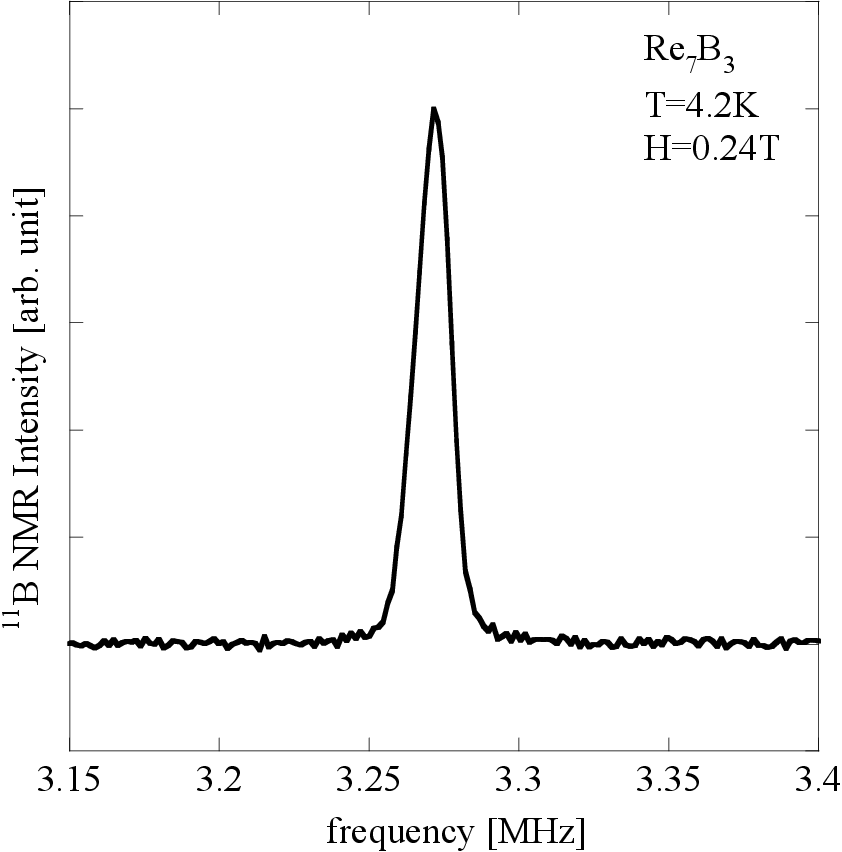}\end{center}
\caption{$^{11}$B NMR spectrum of Re$_7$B$_3$ measured at $T$=4.2 K under a magnetic field of $H$=0.24 T}
\label{f1}
\end{figure}
\subsection{Re$_7$B$_3$}

Figure \ref{f1} shows the $^{11}$B-NMR spectrum at $T$=4.2 K
obtained by fast Fourier transform (FFT) of a spin echo.
The full width at the half maximum (FWHM) of the NMR line is 11.9 kHz.
The very sharp transition indicates the high quality of the sample.

\begin{figure}[htbp]
\begin{center}\includegraphics[clip,width=70mm]{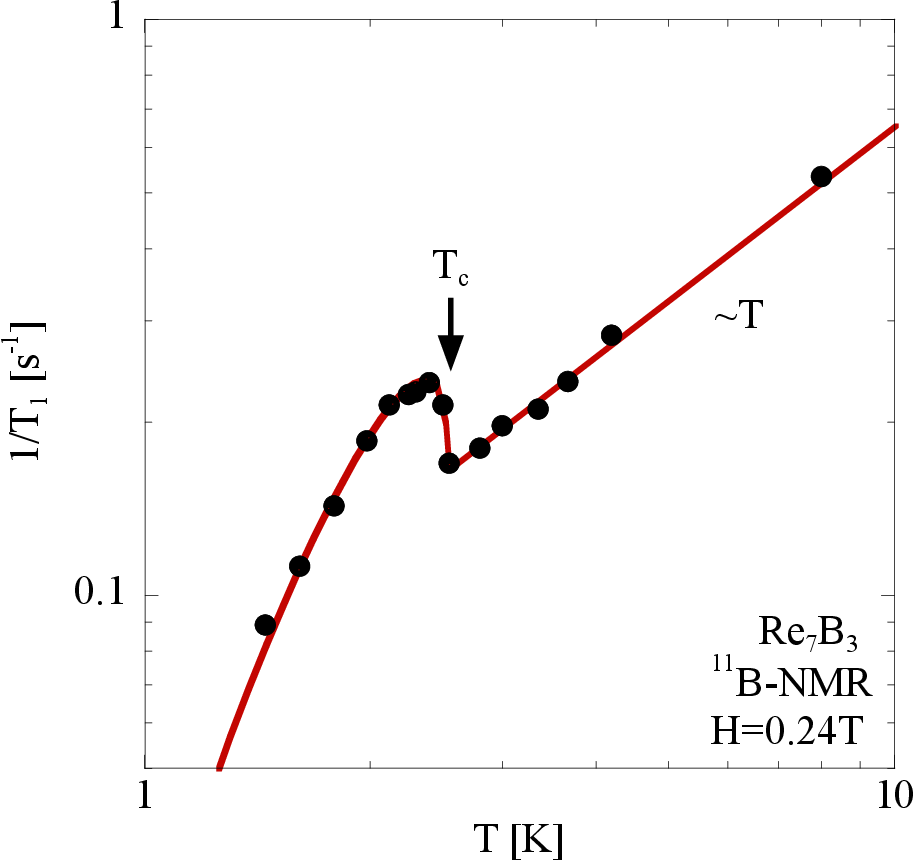}\end{center}
\caption{(color online) 
The temperature dependence of the spin
lattice relaxation rate, \tl\ for Re$_7$B$_3$, measured at a field of $H$=0.24 T.
The straight line above \tc\ represents the $T_1T$=const relation. 
The solid curve below \tc\ is a calculation assuming the BCS gap function. For details, see the text.
}
\label{f2}
\end{figure}

Figure \ref{f2} shows the temperature dependence of \tl\ of $^{11}$B.
The nuclear magnetization decay curve is  fitted to the theoretical formula
\begin{eqnarray}
\frac{M_0-M(t)}{M_0}=\exp\left({-\frac{t}{T_1}}\right),
\end{eqnarray}
with a unique $T_1$ component.
In the normal state above \tc, \tl varies in proportion to $T$, as seen in conventional metals, indicating no electron-electron interaction.
As can be seen clearly in the figure, \tl\ is enhanced just below $T_c$ over its normal-state value, forming
a so-called coherence peak (Hebel-Slichter peak), which is a hallmark of an isotropic superconducting gap.
The $1/T_{1S}$ in the superconducting state is expressed as
\begin{eqnarray}
\frac{T_{1N}}{T_{1S}}=\frac{2}{K_BT} 
\iint\left(1+\frac{\Delta^2}{EE'}\right)N_S(E)N_S(E')  \nonumber \\
\times f(E)\left[1-f(E')\right]\delta(E-E')dEdE',
\end{eqnarray}

where $1/T_{1N}$ is the relaxation rate in the normal state, $N_S(E)$ is the superconducting density of
states (DOS), $f(E)$ is the Fermi distribution function and $C =1+ \frac{\Delta^2}{EE'}$ is the coherence factor. 
Following Hebel, we convolute $N_S(E)$ with a broadening function $B(E)$,
which is approximated with a rectangular function centered at $E$ with a height of 1/2$\delta$.
The solid curve below \tc\ shown in Fig. 2  is a calculation with $2\Delta = 3.2$\kb, $r\equiv\Delta(0)/\delta=4$.
It fits the experimental data reasonably well. The parameter $2\Delta$ is close to the BCS value of 3.5\kb.
This result indicates an isotropic superconducting gap in this compound.

\subsection{LaBiPt}
\begin{figure}[htbp]
\begin{center}\includegraphics[clip,width=70mm]{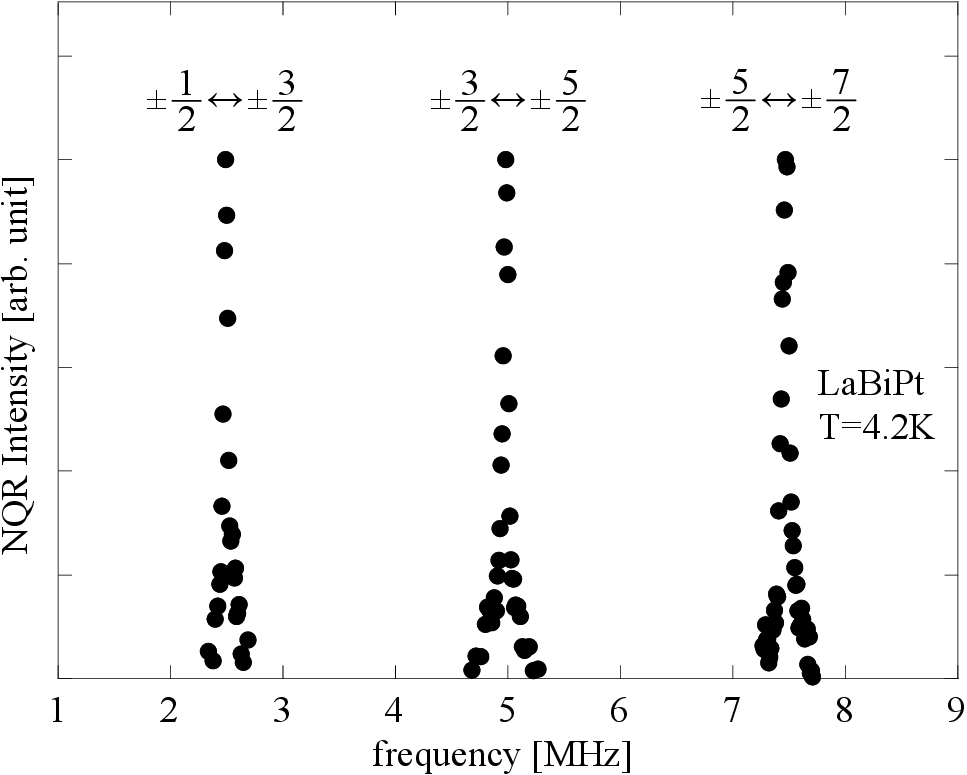}\end{center}
\caption{$^{139}$La NQR spectra of LaBiPt measured at 4.2K.}
\label{f3}
\end{figure}

\begin{figure}[htbp]
\begin{center}\includegraphics[clip,width=70mm]{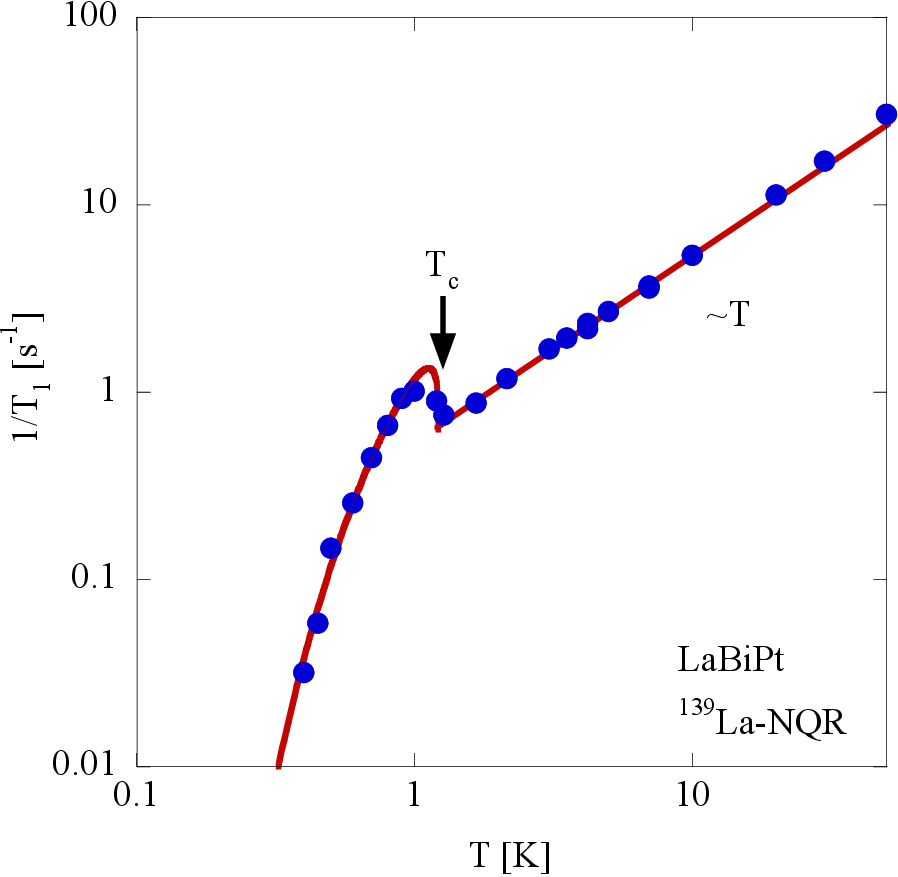}\end{center}
\caption{(color online) The temperature dependence of the $^{139}$La spin-lattice
relaxation rate, \tl measured at the $\pm1/2\leftrightarrow \pm3/2$ NQR transition.
The straight line above \tc\ represents the $T_1T$=const relation. 
The solid curve below \tc\ is a calculation assuming the BCS gap function. For details, see the text.}
\label{f4}
\end{figure}
Figure \ref{f3} shows the $^{139}$La ($I=7/2$) NQR spectra of LaBiPt at $T$=4.2 K. 
Three transition lines centered at 2.49, 4.98 and 7.47 MHz are observed which correspond to the transitions between the
adjacent levels ($\pm1/2\leftrightarrow\pm3/2$), ($\pm3/2\leftrightarrow\pm5/2$) and ($\pm5/2\leftrightarrow\pm7/2$), respectively.
The NQR frequency $\nu_Q = 2.49$ MHz and the asymmetry parameter $\eta = 0$.
Here $\nu_Q$ and $\eta$ are defined as
\begin{eqnarray}
\nu_Q \equiv \nu_z =\frac{3}{2I(2I-1)h}e^2Q\frac{\partial^2 V}{\partial z^2}\\
\eta = \frac{|\nu_x-\nu_y|}{\nu_z},
\end{eqnarray}
with $Q$ being the nuclear quadrupole moment and $\frac{\partial^2 V}{\partial \alpha^2} (\alpha = x, y, z) $
being the electric field gradient at the position of the nucleus.
The FWHM of the NQR line are 28.0 kHz, 37.4 kHz and 50.5kHz for
($\pm1/2\leftrightarrow\pm3/2$), ($\pm3/2\leftrightarrow\pm5/2$) and ($\pm5/2\leftrightarrow\pm7/2$), respectively.
The very sharp transition testifies the high quality of the sample.

Figure \ref{f4} shows the temperature dependence of \tl of $^{139}$La NQR. 
The \tl\ was measured at the 1$\nu_Q$ ($\pm1/2\leftrightarrow\pm3/2$) transition.
The nuclear magnetization decay curve is  fitted to the theoretical formula\cite{eta_not_0}
\begin{eqnarray}
\frac{M_0-M(t)}{M_0}=
0.024\exp\left({-\frac{3t}{T_1}}\right)+
0.234\exp\left({-\frac{10t}{T_1}}\right)\nonumber\\
+0.742\exp\left({-\frac{21t}{T_1}}\right),
\end{eqnarray}
with a unique $T_1$ component.
In the normal state above \tc, \tl varies in proportion to $T$, which indicates no electron-electron interaction.
As can be seen clearly in the figure, \tl\ is enhanced just below $T_c$ over its normal-state value, 
forming a well-defined coherence peak.
The solid curve below \tc\  shown in Fig. 4 is a calculation with $2\Delta = 3.42$\kb\  and $r=20$.
The parameter $2\Delta$ is close to the BCS value of 3.5\kb.
This result also indicates an isotropic superconducting gap.

\subsection{BiPd}

\begin{figure}[htbp]
\begin{center}\includegraphics[clip,width=70mm]{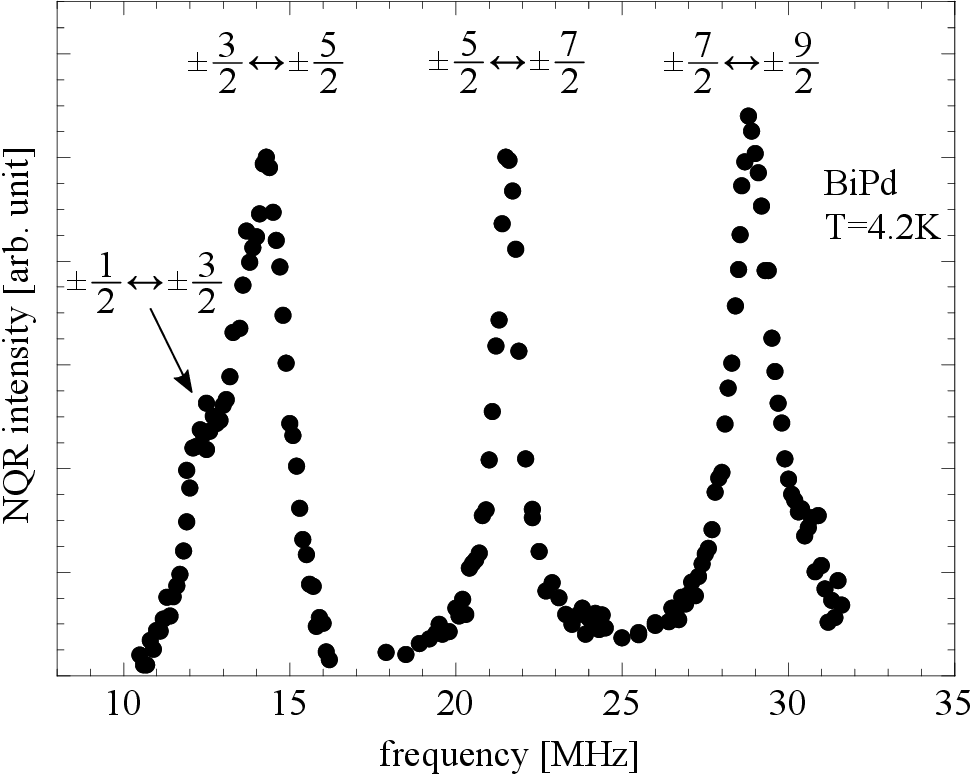}\end{center}
\caption{$^{209}$Bi NQR spectra of BiPd measured at 4.2K.}
\label{f5}
\end{figure}

\begin{figure}[htbp]
\begin{center}\includegraphics[clip,width=70mm]{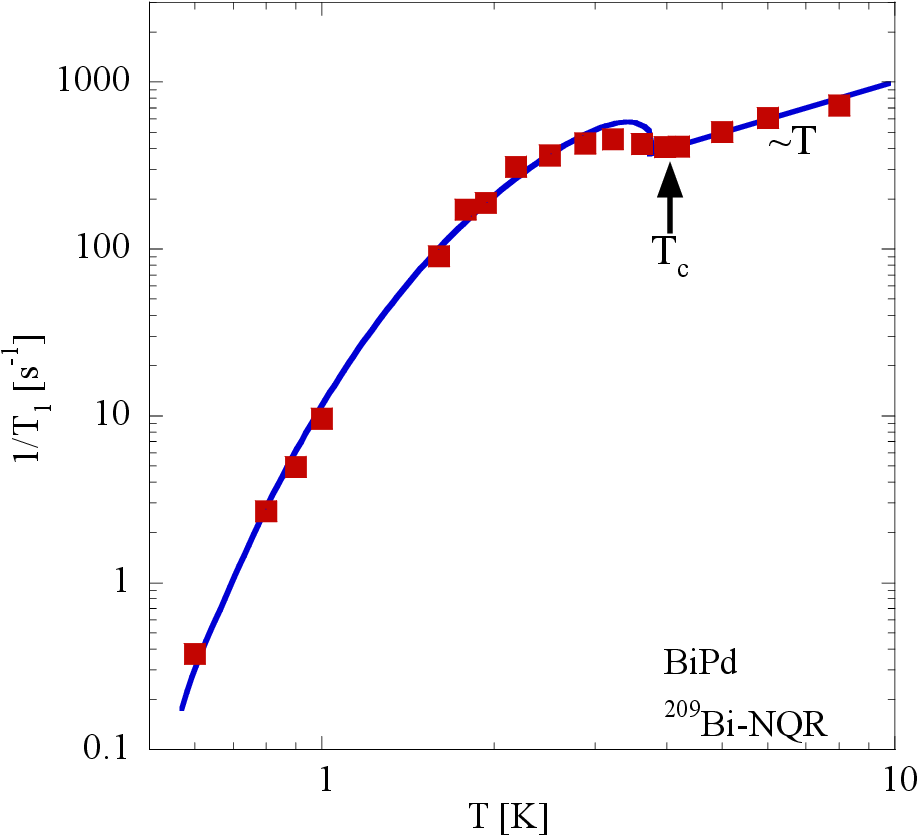}\end{center}
\caption{(color online)
The temperature dependence of the spin lattice relaxation rate, 
\tl\ for BiPd, measured at the $\pm7/2\leftrightarrow\pm9/2$ transition.
The straight line above \tc\ represents the $T_1T$=const relation. 
The solid curve below \tc\ is a calculation assuming the BCS gap function. 
}
\label{f6}
\end{figure}
Figure \ref{f5} shows the $^{209}$Bi ($I=9/2$) NQR spectra of BiPd measured at $T$=4.2 K. 
Four transition lines centered at 12.4, 14.3, 21.4 and 28.8 MHz are observed, which correspond to the transitions between the
adjacent levels ($\pm1/2\leftrightarrow\pm3/2$), ($\pm3/2\leftrightarrow\pm5/2$), ($\pm5/2\leftrightarrow\pm7/2$) and
($\pm7/2\leftrightarrow\pm9/2$), respectively.
The NQR frequency $\nu_Q = 7.31$MHz and the asymmetry parameter $\eta = 0.35$.

Figure \ref{f6} shows the temperature dependence of \tl of $^{209}$Bi,
which was measured at the 4$\nu_Q$ ($\pm7/2\leftrightarrow\pm9/2$) transition.
The nuclear magnetization decay curve is  fitted to the theoretical formula\cite{eta_not_0}
\begin{eqnarray}
\frac{M_0-M(t)}{M_0}=
0.117\exp\left({-\frac{3t}{T_1}}\right)+
0.415\exp\left({-\frac{9t}{T_1}}\right)\nonumber\\
+0.394\exp\left({-\frac{16.5t}{T_1}}\right)+
0.074\exp\left({-\frac{30.5t}{T_1}}\right),
\end{eqnarray}
with a unique $T_1$ component.
In the normal state above \tc, \tl varies in proportion to $T$, as seen in Re$_7$B$_3$ and LaBiPt.
The \tl\ is enhanced just below $T_c$ over its normal-state value, forming
a suppressed, broad coherence peak. The result also indicates an isotropic superconducting gap dominating in BiPd.
The solid curve below \tc\ shown in Fig. \ref{f6}  is a calculation with $2\Delta = 2.7$\kb\ and $r=7$.
The parameter $2\Delta$ is slightly smaller than the BCS value of 3.5\kb.
Previously,  a second superconducting gap was reported to appear below 2 K from the point-contact Andreev reflection measurements.\cite{PhysRevB.86.094520}
However, this feature is not seen in our measurement. 
In a multigap system, \tl\ should show an anomaly around the temperature where the second gap becomes appreciable compared to $k_BT$.\cite{Matano_PrFeAs} 
The absence of such $T_1$ anomaly   suggests that the second gap is small, if any.

\subsection{Discussion}

From the NMR/NQR measurements, the superconducting gap function is found to be  dominantly isotropic in Re$_7$B$_3$, LaBiPt, and BiPd,  even though all these three compounds contain heavy elements. Below we discuss briefly about the difference between these compounds and known  non-centrosymmetric superconductors.
In Re$_7$B$_3$, Re atoms occupy 70\% of the unit cell, 
which means that the heterogeneity of the electric charge density is small. This is probably a possible cause for the small ASOC. In fact, 
in the case of Mg$_{10+x}$Ir$_{19-y}$B,  the heterogeneity caused by defects of Ir was found to enhance ASOC,  and as a result, the spin-triplet component is enhanced.\cite{Tahara_PhysRevB.80.060503}
In LaBiPt, the band splitting by ASOC at the Fermi level is only a few meV,\cite{PhysRevB.63.125115}  which is much smaller than 200 meV of Li$_2$Pt$_3$B.\cite{PhysRevB.72.174505}\cite{Harada_PhysRevB.86.220502}
In BiPd, although all atoms break the inversion symmetry along all direction,
Pd probably makes the main contribution to  the bands near the Fermi level, so that the ASOC is small.
Indeed, the band splitting at the Fermi level was calculated to be around 50 meV \cite{BiPd_JPs} ,
which is smaller than the value of Li$_2$Pt$_3$B, but close to the value of Li$_2$Pd$_3$B (30 meV).\cite{PhysRevB.72.174505}
Our results indicate that not only heavy elements, but also the extent of the inversion symmetry breaking is important.
This is consistent with the finding in  Li$_2$(Pt$_{1-x}$Pd$_x$)$_3$B,@where an abrupt decrease in the Pt(Pd)$_6$B octahedron-octahedron angle was found to dramatically enhance the ASOC.\cite{Harada_PhysRevB.86.220502}

\begin{figure}[htbp]
\begin{center}\includegraphics[clip,width=70mm]{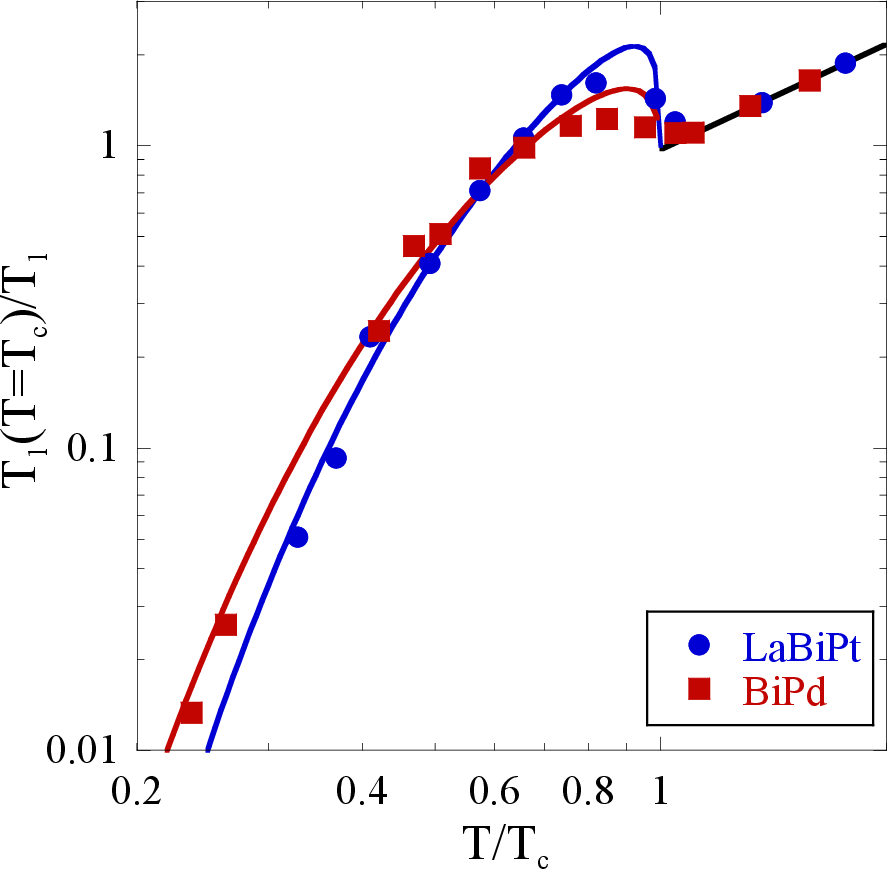}\end{center}
\caption{(color online) Normalized \tl\ against the reduced temperature. The straight line above \tc\ indicates the $T_1T=$const relation. The curves below \tc\  are the same as in Fig. 4 and Fig. 6}
\label{f7}
\end{figure}
Nonetheless, we found a clear correlation between the band splitting caused by the ASOC and the height of the coherence peak. 
Figure 7 shows the  \tl\ normalized by its value at $T_c$ against the reduced temperature for LaBiPt and BiPd.
Generally, the height of the coherence peak can be affected by many factors such as the anisotropy of superconducting gap,
 impurity scattering, and phonon scattering.\cite{MacLaughlin19761}
Here, we propose an alternative possibility.
When a spin-triplet component is induced, the corresponding gap component will have nodes. 
As a result, the observed height of the coherence peak will be suppressed.  
The coherence peak for BiPd is much smaller than that of LaBiPt. 
As mentioned already, the band splitting of BiPd (50 meV) is much larger than that of LaBiPt (a few meV). 
A similar result was previously observed in Mg$_{10+x}$Ir$_{19-y}$B, where the coherence peak is suppressed when a spin-triplet component is enhanced as evidenced by the Knight shift measurement.\cite{Tahara_PhysRevB.80.060503}
If it is also the case here, our result suggests that there exists substantial component of the spin-triplet state in BiPd. The Knight shift measurement in a single crystal sample is required to quantitatively determine such component.

\section{Summary}
We have presented  NMR/NQR measurements on the non-centrosymmetric superconductors Re$_7$B$_3$, LaBiPt, and BiPd containing heavy elements.
The spin-lattice relaxation rate \tl\ shows a coherence peak  below \tc\ and follows an exponential $T$ variation at low temperatures for all samples.
This result indicate that an isotropic superconducting gap is dominant in Re$_7$B$_3$, LaBiPt, and BiPd. However, the much suppressed coherence peak height in BiPd suggests  that  a substantial component of the spin-triplet state with nodes in the gap function exists in BiPd. 
Our results indicate that   heavy element is not the only important factor for  unconventional superconductivity to appear, but the crystal structure that breaks extensively an inversion symmetry is required.
 
\begin{acknowledgment}
We acknowledge partial support by the ``Topological
Quantum Phenomena" Grant-in Aid for Scientific Research on innovative Areas from MEXT of Japan (Grant No. 22103004) and JSPS grant No.20244058.
\end{acknowledgment}



\begin{thebibliography}{10}

\bibitem{PhysRevLett.87.037004}
L.~P. Gor'kov and E.~I. Rashba: Phys. Rev. Lett. {\bfseries 87} (2001) 037004.

\bibitem{PhysRevLett.92.097001}
P.~A. Frigeri, D.~F. Agterberg, A.~Koga, and M.~Sigrist: Phys. Rev. Lett.
  {\bfseries 92} (2004) 097001.

\bibitem{1367-2630-6-1-115}
P.~A. Frigeri, D.~F. Agterberg, and M.~Sigrist: New Journal of Physics
  {\bfseries 6} (2004) 115.

\bibitem{PhysRevLett.105.097002}
Y.~Tanaka, Y.~Mizuno, T.~Yokoyama, K.~Yada, and M.~Sato: Phys. Rev. Lett.
  {\bfseries 105} (2010) 097002.

\bibitem{RevModPhys.83.1057}
X.-L. Qi and S.-C. Zhang: Rev. Mod. Phys. {\bfseries 83} (2011) 1057.

\bibitem{PhysRevLett.92.027003}
E.~Bauer, G.~Hilscher, H.~Michor, C.~Paul, E.~W. Scheidt, A.~Gribanov,
  Y.~Seropegin, H.~No\"el, M.~Sigrist, and P.~Rogl: Phys. Rev. Lett. {\bfseries
  92} (2004) 027003.

\bibitem{UIr}
T.~Akazawa, H.~Hidaka, T.~Fujiwara, T.~C. Kobayashi, E.~Yamamoto, Y.~Haga,
  R.~Settai, and Y.~Onuki: Journal of Physics: Condensed Matter {\bfseries 16}
  (2004) L29.

\bibitem{CeRhSi3}
N.~Kimura, K.~Ito, K.~Saitoh, Y.~Umeda, H.~Aoki, and T.~Terashima: Phys. Rev.
  Lett. {\bfseries 95} (2005) 247004.

\bibitem{CeIrSi3}
I.~Sugitani, Y.~Okuda, H.~Shishido, T.~Yamada, A.~Thamizhavel, E.~Yamamoto,
  T.~D. Matsuda, Y.~Haga, T.~Takeuchi, R.~Settai, and Y.~\={O}nuki: Journal of
  the Physical Society of Japan {\bfseries 75} (2006) 043703.

\bibitem{Li2Pd3B}
K.~Togano, P.~Badica, Y.~Nakamori, S.~Orimo, H.~Takeya, and K.~Hirata: Phys.
  Rev. Lett. {\bfseries 93} (2004) 247004.

\bibitem{Li2Pt3B}
P.~Badica, T.~Kondo, and K.~Togano: J. Phys. Soc. Jpn. {\bfseries 74} (2005)
  1014.

\bibitem{Mg10Ir19B16}
T.~Klimczuk, Q.~Xu, E.~Morosan, J.~D. Thompson, H.~W. Zandbergen, and R.~J.
  Cava: Phys. Rev. B {\bfseries 74} (2006) 220502.

\bibitem{Nishiyama_PhysRevB.71.220505}
M.~Nishiyama, Y.~Inada, and G.-q. Zheng: Phys. Rev. B {\bfseries 71} (2005)
  220505.

\bibitem{Nishiyama_PhysRevLett.98.047002}
M.~Nishiyama, Y.~Inada, and G.-q. Zheng: Phys. Rev. Lett. {\bfseries 98} (2007)
  047002.

\bibitem{Harada_PhysRevB.86.220502}
S.~Harada, J.~J. Zhou, Y.~G. Yao, Y.~Inada, and G.-q. Zheng: Phys. Rev. B
  {\bfseries 86} (2012) 220502.

\bibitem{Re7B3}
A.~Kawano, Y.~Mizuta, H.~Takagiwa, T.~Muranaka, and J.~Akimitsu: J. Phys. Soc.
  Jpn. {\bfseries 72} (2003) 1724.

\bibitem{LaBiPt}
G.~Goll, M.~Marz, A.~Hamann, T.~Tomanic, K.~Grube, T.~Yoshino, and
  T.~Takabatake: Physica B: Condensed Matter {\bfseries 403} (2008) 1065 .

\bibitem{nature-mat-9-546}
H.~Lin, L.~A. Wray, Y.~Xia, S.~Xu, S.~Jia, R.~J. Cava, A.~Bansil, and M.~Z.
  Hasan: Nat. Mater. {\bfseries 9} (2010) 546.

\bibitem{PhysRevLett.105.096404}
D.~Xiao, Y.~Yao, W.~Feng, J.~Wen, W.~Zhu, X.-Q. Chen, G.~M. Stocks, and
  Z.~Zhang: Phys. Rev. Lett. {\bfseries 105} (2010) 096404.

\bibitem{PhysRevB.84.220504}
N.~P. Butch, P.~Syers, K.~Kirshenbaum, A.~P. Hope, and J.~Paglione: Phys. Rev.
  B {\bfseries 84} (2011) 220504.

\bibitem{2013arXiv1302.1943T}
F.~F. {Tafti}, T.~{Fujii}, A.~{Juneau-Fecteau}, S.~R. {de Cotret},
  N.~{Doiron-Leyraud}, A.~{Asamitsu}, and L.~{Taillefer}: Phys. Rev. B {\bfseries
  87} (2013) 184504.

\bibitem{BiPd}
B.~Joshi, A.~Thamizhavel, and S.~Ramakrishnan: Phys. Rev. B {\bfseries 84}
  (2011) 064518.

\bibitem{PhysRevB.86.094520}
M.~Mondal, B.~Joshi, S.~Kumar, A.~Kamlapure, S.~C. Ganguli, A.~Thamizhavel,
  S.~S. Mandal, S.~Ramakrishnan, and P.~Raychaudhuri: Phys. Rev. B {\bfseries
  86} (2012) 094520.

\bibitem{eta_not_0}
J.~Chepin and J.~H.~R. Jr: Journal of Physics: Condensed Matter {\bfseries 3}
  (1991) 8103.

\bibitem{Matano_PrFeAs}
K.~Matano, Z.~A. Ren, X.~L. Dong, L.~L. Sun, Z.~X. Zhao, and G.~qing Zheng: EPL
  (Europhysics Letters) {\bfseries 83} (2008) 57001.

\bibitem{Tahara_PhysRevB.80.060503}
K.~Tahara, Z.~Li, H.~X. Yang, J.~L. Luo, S.~Kawasaki, and G.-q. Zheng: Phys.
  Rev. B {\bfseries 80} (2009) 060503.

\bibitem{PhysRevB.63.125115}
T.~Oguchi: Phys. Rev. B {\bfseries 63} (2001) 125115.

\bibitem{PhysRevB.72.174505}
K.-W. Lee and W.~E. Pickett: Phys. Rev. B {\bfseries 72} (2005) 174505.

\bibitem{BiPd_JPs}
K.~Okawa, M.~Kanou, T.~Katagiri, H.~Kashiwaya, S.~Kashiwaya, and T.~Sasagawa:
 presented at JPS Meet., March 2013, Hiroshima, 29aXJ-3 .

\bibitem{MacLaughlin19761}
D.~E. MacLaughlin: in {\em Magnetic Resonance in the Superconducting State}, ed.
  F.~S. Henry~Ehrenreich and D.~Turnbull (Academic Press, 1976), Vol.~31 of
  {\em Solid State Physics}, pp. 1 -- 69.

\end{thebibliography}
\end{document}